\documentclass[a4paper, 11pt]{article}

\usepackage{amsmath}
\usepackage{amsfonts}
\usepackage{amssymb}
\usepackage{amsxtra}
\usepackage{floatflt}
\usepackage{comment}
\usepackage[pdftex]{graphicx}
\usepackage{enumerate}
\usepackage{float}
\usepackage[normalem]{ulem}
\usepackage{color}
\usepackage{mathtools}
\graphicspath{{./figs/}}
\usepackage{marvosym}
\usepackage{times}
\usepackage{bm}
\usepackage{colortbl} 
\usepackage{multirow}
\usepackage{graphicx}
\usepackage{nameref}
\usepackage{xcolor}

\usepackage[usestackEOL]{stackengine}

\usepackage{tabularx}
\newcolumntype{L}[1]{>{\raggedright\arraybackslash}p{#1}} 
\newcolumntype{C}[1]{>{\centering\arraybackslash}p{#1}} 
\newcolumntype{R}[1]{>{\raggedleft\arraybackslash}p{#1}}

\newcommand{\ly}{\left\{}
\newcommand{\ry}{\right\}}

\newcommand{\eg}{e.g.~}
\newcommand{\ie}{i.e.~}

\newcommand{\figref}[1]{Fig.~\ref{#1}}

\usepackage{textcomp}
\usepackage{xcolor,colortbl}
\usepackage{enumitem}

\newcommand{\e}[1]{_{\text{#1}}}

\definecolor{Gray}{gray}{0.95}
\definecolor{LightCyan}{rgb}{0.9,1,1}
\newcolumntype{a}{>{\columncolor{LightCyan}}c}
\newcolumntype{b}{>{\columncolor{white}}c}

\usepackage{fancyhdr}
\begin{document}

\noindent
\LARGE\textbf{Prediction of cellular burden with host-circuit models} \vspace{2mm}\\
\large \textbf{Evangelos-Marios Nikolados}$^{1}$, \textbf{Andrea Y. Wei\ss e}$^{2}$, \textbf{Diego A. Oyarz\'un}$^{1,3}$ \\
\normalsize{$^{1}$ School of Biological Sciences, University of Edinburgh, UK; e.m.nikolados@sms.ed.ac.uk} \\
\normalsize{$^{2}$ Department of Infectious Diseases, Imperial College London, UK; andrea.weisse@imperial.ac.uk} \\
\normalsize{$^{3}$ School of Informatics, University of Edinburgh, UK; d.oyarzun@ed.ac.uk}

\paragraph{Summary} 
Heterologous gene expression draws resources from host cells. These resources include vital components to sustain growth and replication, and the resulting \emph{cellular burden} is a widely recognised bottleneck in the design of robust circuits. 
In this tutorial we discuss the use of computational models that integrate gene circuits and the physiology of host cells. Through various use cases, we illustrate the power of host-circuit models to predict the impact of design parameters on both burden and circuit functionality. 
Our approach relies on a new generation of computational models for microbial growth that can flexibly accommodate resource bottlenecks encountered in gene circuit design. Adoption of this modelling paradigm can facilitate fast and robust design cycles in synthetic biology. 

\section{Introduction}
The grand goal of Synthetic Biology is to engineer living systems with novel functions. The approach relies on the combination of biological knowledge with design strategies from engineering sciences \cite{andrianantoandro2006synthetic, canton2008refinement, ninfa2007using, teo2015synthetic}. Engineering principles, such as modularity and standardisation, have led to gene circuits with a wide range of functions such as cellular oscillators \cite{elowitz2000synthetic,hasty2002engineered}, memory devices \cite{gardner2000construction} and biosensors \cite{tabor2009synthetic,Mannan2017}. As synthetic biology matures into an engineering discipline of its own, mathematical modelling is playing an increasingly important role in the design of biological circuitry \cite{Oyarzun2013}. Moreover, model-based design offers opportunities for other fields such as computer-aided design \cite{nielsen2016genetic}, control theory \cite{Chaves2019} and machine learning \cite{Carbonell2019} to contribute with new methods and protocols for gene circuit design.

The success of the celebrated ``design-build-test-learn'' cycle \cite{Hughes2017} relies on the availability of good quality models for circuit function. A major drawback of current modelling frameworks, however, is the implicit assumption that biological circuits function in isolation from their host. This simplification limits the predictive power of circuit models and slows down the iterations between system design, testing and characterisation. In reality, gene circuits interact with their host in many ways, including the consumption of molecular resources such as amino acids, nucleotides or energy, as well as using major components of the genetic machinery such as polymerases and ribosomes. 

Competition for a limited pool of host resources produces a two-way interplay between synthetic circuits and the native physiology of the host \cite{rondelez2012competition}. This interplay is commonly known as \textit{burden} and perturbs the homeostatic balance of the host, resulting in slowed growth, reduced biosynthesis and the induction of stress responses \cite{cardinale2012contextualizing}. Since such effects can impact circuit behaviour, they create feedback effects that can potentially break down circuit function \cite{gyorgy2014limitations, mather2013translational, scott2010interdependence}. As a result, individual modelling of circuit parts and their connectivity is not sufficient to predict circuit function accurately.

In a seminal study on host-circuit interactions, Tan and colleagues \cite{tan2009emergent} studied a simple circuit consisting of T7 RNA polymerase that activates its own expression in \emph{Escherichia coli}. Contrary to what standard mathematical models would predict, the circuit displayed bistable dynamics. The authors show that synthesis of the polymerase produced an indirect, growth-mediated, positive feedback loop, which when included in their model was able to reproduce the observed bistability. This study was the first empirical demonstration that growth defects can drasticlly change circuit function. A number of subsequent works have focused on the sources and impact of burden on gene circuits. For example, Ceroni \emph{et al} showed that genes with weaker ribosomal binding strength are less taxing on the host resources \cite{ceroni2015quantifying}. Other works have focused on strategies to mitigate burden. An and Chin built a gene expression system that combines orthogonal transcription by T7 RNA polymerase and translation by orthogonal ribosomes \cite{an2009synthesis}. The system reported in \cite{segall2014resource} allows to allocate resources among competing genes, while \cite{pasini2016using} built libraries of promoters that tune expression of burdensome proteins and decrease cellular stress. The work by Shopera \emph{et al} showed that negative feedback control can reduce the cross-talk between gene circuits \cite{shopera2017decoupling}. Another strategy for reducing burden was proposed in \cite{Darlington2018} using an orthogonal ribosome for translation of heterologous genes. A particularly attractive strategy is to exploit burden to improve functionality. For example, Rugbjerg and colleagues increased metabolite production by coupling pathway expression to that of essential endogenous genes \cite{rugbjerg2018synthetic}, while \cite{ceroni2018burden} employed stress-response promoters to build a feedback system with increased protein yield.

As a result of the increasing interest in cellular burden and host-circuit interactions, the modelling community has devoted substantial attention to improving models for gene circuits and their interaction with a host. A key challenge is to find a suitable level of model complexity with enough detail to describe tunable circuit parts but without excessive granularity that makes models impractical. At one end of the complexity spectrum, a number of works have proposed simple resource allocation models for the interplay between circuit and host genes \cite{gyorgy2015isocost,carbonell2015dealing,gorochowski2016minimal}. Using different modelling approaches and assumptions, these models generally predict a linear relation between expression of native and heterologous genes. Increases in the expression of one gene causes a linear drop in the expression of another gene, as a result of a limited abundance of ribosomes for translation. At the other end of the spectrum, the whole-cell model of \emph{Mycoplasma genitalium} \cite{karr2012whole} was an ambitious attempt to describe all layers of cellular organization under a single computational model. A subsequent work demonstrated the use of the whole-cell model in conjunction with gene circuits \cite{Purcell2013}. Yet to date such whole-cell models have not been built for bacterial hosts commonly employed in synthetic biology, and their high complexity prevents their systematic use in circuit design and optimization. 

A number of approaches have sought to find a middle ground between model complexity and tractability. Inspired by the widely established ``bacterial growth laws'' \cite{Klumpp2009,scott2010interdependence}, Wei\ss e and colleagues built a mechanistic growth model for \emph{Escherichia coli} \cite{weisse2015mechanistic}. The model uses a coarse-grained partition of the proteome to describe how cells allocate their resources across various gene expression tasks. It accurately predicts growth rate from the interplay between metabolism and gene expression, and can be extended with a wide range of genetic circuits. Applications of the Wei\ss e model include the design of orthogonal ribosomes \cite{Darlington2018}, the addition of extra layers of regulation \cite{liao2017integrative} and its extension to single-cell growth dynamics \cite{Thomas2018}. Most recently, Nikolados \emph{et al} employed the model to study the impact of growth defects in various exemplar circuits \cite{nikolados2019growth}. 

\begin{figure}[H]
\centering
\includegraphics[scale=1.3]{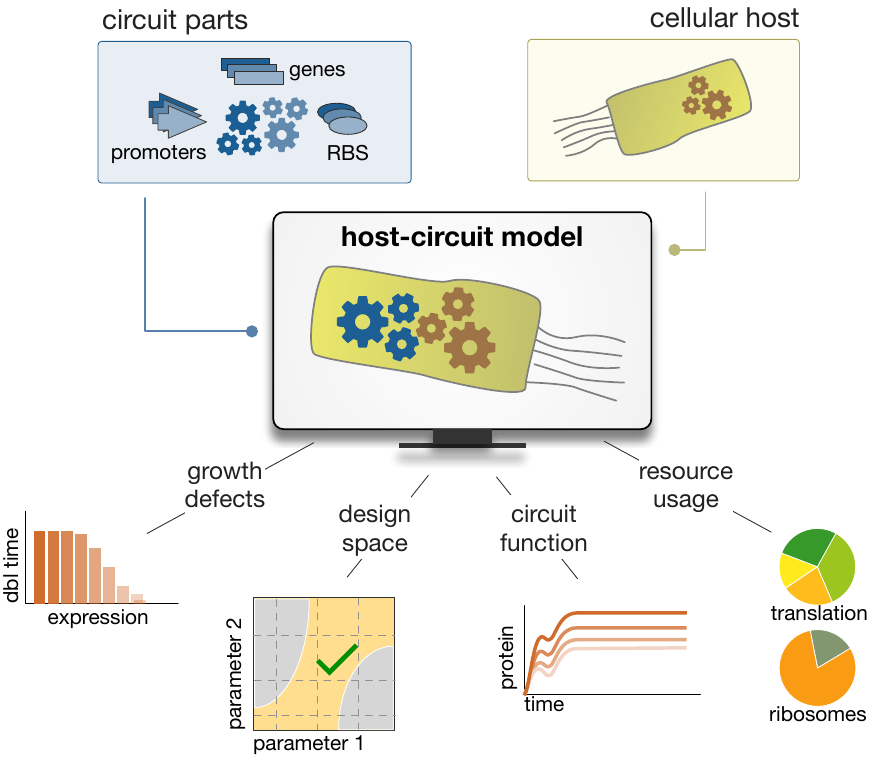}
\caption{\textbf{Host-circuit modelling.} Integrated host-circuit models provide a quantitative basis to study the impact of design parameters on circuit function and genetic burden on their host.}
\label{Fig1}
\end{figure}

In this tutorial we describe how mechanistic growth models can be employed to simulate gene circuits together with the host physiology (\figref{Fig1}). In Section \ref{sec:models} we first revisit the bacterial growth laws and explain the core principles of the mechanistic growth model. In Section \ref{sec:host-circuit} we present how to extend the growth model with heterologous genes. We illustrate the methodology with a number of transcriptional logic gates in Section \ref{sec:gates}. We conclude the chapter with a perspective for future research in the field.

\section{Coarse-grained models for bacterial growth}\label{sec:models}
We begin by describing the bacterial growth laws that form the basis for most current models for growth. Our focus is on \emph{coarse-grained} models that describe cell physiology using lumped variables representing aggregates of molecular species. We deliberately exclude whole-cell models \cite{karr2012whole} and genome-scale models \cite{OBrien2013}, both of which have been discussed extensively in the literature \cite{Carrera2015,Karr2015,OBrien2015} and so far have found relatively limited applications in gene circuit design.

\subsection{Bacterial growth laws}\label{sec:growthlaws}

Bacterial growth has been an active topic of study for many decades. The celebrated work of Nobel laureate Jacques Monod provided a key quantitative description for growth \cite{monod1949growth}, based on the observation that bacteria in batch cultures exhibit several phases of growth:
\begin{itemize}[noitemsep]
    \item \textbf{Lag phase:} cells do not immediately start to grow after nutrient induction, as they first must adapt to the new environment; RNA and proteins are produced as the cell prepares for division.
    \item \textbf{Exponential phase:} cells duplicate at a constant rate, so that their number grows exponentially as $N(t) = N\e{0}2^{t/\tau}$ with $\tau$ being the average doubling time. Equivalently, the number of cells can be expressed as $N(t) = N\e{0}e^{\lambda t}$, where $\lambda=\log 2/\tau$ is the growth rate. 
    \item \textbf{Stationary phase:} cell replication stops because an essential nutrient has been depleted from the batch. The number of cells remains constant during this phase.
    \item \textbf{Death phase:} cells begin to die, resulting in a decreasing cell population.
\end{itemize}

The vast majority of studies on bacterial growth focus on the exponential phase, and to date this remains the best characterised growth phase. A widely empirical model for exponential growth is given by Monod's law, which relates the instantaneous growth rate and the substrate concentration:
\begin{equation}
    \lambda = \frac{\lambda \e{max}s}{s + K\e{s}}, \label{eq:1stlaw}
\end{equation}
where $s$ is the growth substrate, $\lambda \e{max}$ is the maximum growth rate possible in the substrate and $K\e{s}$ is the substrate concentration for which growth rate is half maximal. The relationship in Eq.~\eqref{eq:1stlaw} is known as \textit{Monod's law} and describes the hyperbolic dependence of the growth rate $\lambda$ on the concentration of a growth-limiting nutrient $s$ in the medium.

Measurements of bacterial cells growing at different rates \cite{schaechter1958dependency,neidhardt1960studies} have revealed a central role of ribosome synthesis in maintaining exponential growth \cite{dennis2004control,maaloe1979regulation}. In particular, the ribosomal mass fraction, $\phi\e{R}$, has been shown to increase linearly with growth rate
\cite{bremer1996modulation,schaechter1958dependency}. This is the second growth law,  described mathematically as:
\begin{align}
\phi\e{R} = \phi\e{R}^{min} + \frac{\lambda}{\kappa\e{t}},\label{eq:2ndlaw}
\end{align}
where $\phi\e{R}^{\text{min}}$ is an offset term and $\kappa\e{t}$ is a phenomenological parameter related to protein synthesis. 

The third growth law relates to growth inhibition. It has been shown that sublethal antibiotic doses targeting ribosomal activity produce a negative linear relation between growth rate and the ribosomal mass fraction \cite{scott2010interdependence}. Mathematically, this growth law can be described by:
\begin{align}
\phi\e{R} = \phi\e{R}^{\text{max}} - \frac{\lambda}{\kappa\e{n}},\label{eq:3rdlaw}
\end{align}
where the parameter $\kappa\e{n}$ describes the nutrient capacity of the growth medium and $\phi\e{R}^{\text{max}}$ is the maximum allocation to ribosomal synthesis in the limit of complete translational inhibition. 

Taken together, Equations~\eqref{eq:1stlaw}--\eqref{eq:3rdlaw} provide a remarkably simple description of exponential growth. Yet a common caveat of such descriptions is their lack of explicit links between phenomenological parameters and the molecular processes that drive growth. Some works have indeed found quantitative descriptions of model parameters in terms of intracellular properties \cite{Klumpp2009,scott2010interdependence}. However, another strand of research has moved away from phenomenological models toward mechanistic descriptions of cell physiology \cite{maitra2015bacterial,bosdriesz2015fast}. Notably, earlier work by Molenaar and colleagues \cite{molenaar2009shifts} proposed a model that integrates metabolism and protein biosynthesis into a resource allocation model. Key assumption in that approach is that microbes adjust their proteome composition to maximize growth. This leads to growth predictions that rely on an optimality principle, without the need of a mechanistic description of how cellular constituents contribute to growth and replication.

\subsection{A mechanistic model of bacterial growth}\label{sec:weissemodel}
The mechanistic model in  \cite{weisse2015mechanistic} describes bacterial growth based on first principles. The model reproduces the bacterial growth laws and, at the same time, contains detailed mechanisms for nutrient metabolism, transcription and translation. It employs a partition of the proteome similar to an earlier work \cite{molenaar2009shifts}, but it does not require the assumption of growth maximization. The model is versatile and can predict how cells reallocate their proteome composition under various types of perturbations, including nutrient shifts, genetic modifications and antibiotic treatments.

\begin{figure*}[!h]
\centering
\includegraphics[scale=1.3]{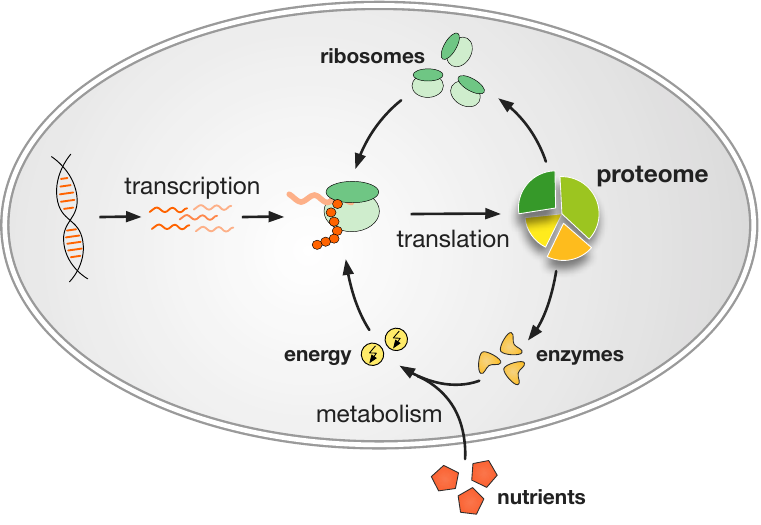}
\caption{\textbf{Mechanistic model for bacterial growth.} The model predicts growth rate from the allocation of two cellular resources (energy and ribosomes) among the various processes that fuel growth and replication \cite{weisse2015mechanistic}.}
\label{Fig2}
\end{figure*}

The model combines nutrient import and its conversion  to cellular energy with the biosynthetic processes of transcription and translation. In its basic form, the model includes 14 intracellular variables: an internalised nutrient \textit{s\textsubscript{i}}; a generic form of energy, denoted \textit{a}, that models the total pool of intracellular molecules required to fuel biosynthesis, such as ATP and aminoacids; and four types of proteins: ribosomes \textit{p\textsubscript{r}}, transporter enzymes \textit{p\textsubscript{t}}, metabolic enzymes \textit{p\textsubscript{m}} and house-keeping proteins \textit{p\textsubscript{q}}. The model also contains the corresponding free and ribosome-bound mRNAs for each protein type, denoted by \textit{m\textsubscript{x}} and \textit{c\textsubscript{x}} respectively, with $x\in\{r,t,m,q\}$. The model can be described by the chemical reactions listed in Table \ref{table:model}. From these reactions we model the cell as a system of ordinary differential equations, describing the rate of change of the numbers of molecules per cell of a particular species. Next we explain in detail how the model equations are built.

\begin{table}[!h]
\caption{Chemical reactions in the mechanistic growth model \cite{weisse2015mechanistic}. \label{table:model}}
\centering
\resizebox{\textwidth}{!}{%
\begin{tabular}{bababac}
 & transcription & dilution/degradation & ribosome binding & dilution & translation & dilution\\

\rowcolor{Gray} ribosomes & \cellcolor{LightCyan} $\phi \xrightarrow{\textit{w\textsubscript{r}}} \textit{m\textsubscript{r}}$ &  $\textit{m\textsubscript{r}} \xrightarrow{\textit{$\lambda +$ d\textsubscript{m}}} \phi$ & \cellcolor{LightCyan} $\textit{p\textsubscript{r}} + \textit{m\textsubscript{r}} \xrightleftharpoons[\textit{k\textsubscript{u}}]{\textit{k\textsubscript{b}}} \textit{c\textsubscript{r}}$ & $\textit{c\textsubscript{r}} \xrightarrow{\textit{$\lambda$}} \phi$ & \cellcolor{LightCyan} $\textit{n\textsubscript{r} a} + \textit{c\textsubscript{r}} \xrightarrow{\textit{v\textsubscript{r}}} \textit{\textit{p\textsubscript{r}} + m\textsubscript{r} + \textit{p\textsubscript{r}}}$ & $\textit{p\textsubscript{r}} \xrightarrow{\textit{$\lambda$}} \phi$\\ 
 
 transporter enzyme 
 & $\phi \xrightarrow{\textit{w\textsubscript{t}}} \textit{m\textsubscript{t}}$ & $\textit{m\textsubscript{t}} \xrightarrow{\textit{$\lambda +$ d\textsubscript{m}}} \phi$ &$\textit{p\textsubscript{r}} + \textit{m\textsubscript{t}} \xrightleftharpoons[\textit{k\textsubscript{u}}]{\textit{k\textsubscript{b}}} \textit{c\textsubscript{t}}$ &$\textit{c\textsubscript{t}} \xrightarrow{\textit{$\lambda$}} \phi$ & $\textit{n\textsubscript{t} a} + \textit{c\textsubscript{t}} \xrightarrow{\textit{v\textsubscript{t}}} \textit{\textit{p\textsubscript{r}} + m\textsubscript{t} + p\textsubscript{t}}$ & $\textit{p\textsubscript{t}} \xrightarrow{\textit{$\lambda$}} \phi$\\ 
 
\rowcolor{Gray} metabolic enzyme &\cellcolor{LightCyan}
  $\phi \xrightarrow{\textit{w\textsubscript{m}}} \textit{m\textsubscript{m}}$ &  $\textit{m\textsubscript{m}} \xrightarrow{\textit{$\lambda +$ d\textsubscript{m}}} \phi$ & \cellcolor{LightCyan} $\textit{p\textsubscript{r}} + \textit{m\textsubscript{m}} \xrightleftharpoons[\textit{k\textsubscript{u}}]{\textit{k\textsubscript{b}}} \textit{c\textsubscript{m}}$ &$\textit{c\textsubscript{m}} \xrightarrow{\textit{$\lambda$}} \phi$  & \cellcolor{LightCyan} $\textit{n\textsubscript{m} a} + \textit{c\textsubscript{m}} \xrightarrow{\textit{v\textsubscript{m}}} \textit{\textit{p\textsubscript{r}} + m\textsubscript{m} + p\textsubscript{m}}$ & $\textit{p\textsubscript{m}} \xrightarrow{\textit{$\lambda$}} \phi$\\ 
 
  house-keeping proteins 
 & $\phi \xrightarrow{\textit{w\textsubscript{q}}} \textit{m\textsubscript{q}}$ & $\textit{m\textsubscript{q}} \xrightarrow{\textit{$\lambda +$ d\textsubscript{m}}} \phi$ &$\textit{p\textsubscript{r}} + \textit{m\textsubscript{q}} \xrightleftharpoons[\textit{k\textsubscript{u}}]{\textit{k\textsubscript{b}}} \textit{c\textsubscript{q}}$ &$\textit{c\textsubscript{q}} \xrightarrow{\textit{$\lambda$}} \phi$ & $\textit{n\textsubscript{q} a} + \textit{c\textsubscript{q}} \xrightarrow{\textit{v\textsubscript{q}}} \textit{\textit{p\textsubscript{r}} + m\textsubscript{q} + \textit{p\textsubscript{q}}}$& $\textit{p\textsubscript{q}}  \xrightarrow{\textit{$\lambda$}} \phi$\\\\

\rowcolor{Gray} nutrient import & \cellcolor{LightCyan} $s \xrightarrow{v\e{imp}} s\e{i}$ & internal nutrient & \cellcolor{LightCyan} $s\e{i} \xrightarrow{\lambda} \phi$\\

 metabolism & \cellcolor{LightCyan} $s\e{i} \xrightarrow{v\e{cat}} n\e{s} a$ &  energy molecules & \cellcolor{LightCyan} $a \xrightarrow{\lambda} \phi$
\end{tabular}}
\end{table}

The environment, or growth medium, of the cell contains a single nutrient described by the constant parameter $s$. A transport protein \textit{p\textsubscript{t}} is responsible for the uptake of the external nutrient at a fixed concentration, which once internalised, \textit{s\textsubscript{i}}, is catabolised by a metabolic enzyme \textit{p\textsubscript{m}}. The dynamics of the internalised nutrient obey:
\begin{align}
    \dot{s}\e{i} &= v\e{imp} - v\e{cat} - \lambda s\e{i}.\label{eq:nutrient}
\end{align}
Similarly to the bacterial growth laws described in Section \ref{sec:growthlaws}, the growth rate is denoted by $\lambda$. All intracellular species are assumed to be diluted at a rate $\lambda$ because of partitioning cellular content between daughter cells at division. Nutrient import ($v\e{imp}$) and catabolism ($v\e{cat}$) are assumed to follow Michaelis-Menten kinetics:
\begin{equation}
    v\e{imp} = p\e{t} \frac{v\e{t}s}{K\e{t} + s}, \quad\quad\quad\quad v\e{cat} = p\e{m} \frac{v\e{m}s\e{i}}{K\e{m} + s\e{i}},
\end{equation}
where $v\e{t}$ and  $v\e{m}$ are maximal rates, while $K\e{t}$ and $K\e{m}$ are  Michaelis-Menten constants. Since translation is known to dominate energy consumption \cite{bremer1996modulation}, the model neglects other energy-consuming processes. Using $c\e{x}$ to denote the complex between a ribosome and the mRNA for a protein $p\e{x}$, the translation rate for every protein obeys
\begin{equation}
    v\e{x} = c\e{x}\frac{\gamma(a)}{n\e{x}}. \label{eq:trrate}
\end{equation}
The parameter $n\e{x}$ in Eq.~\eqref{eq:trrate} is the length of the protein $p\e{x}$ in terms of amino acids, and the term $\gamma(a)$ represents the net rate of translational elongation. Assuming that each elongation step consumes a fixed amount of energy \cite{weisse2015mechanistic}, the net elongation rate depends on the energy resource by:
\begin{align}
    \gamma(a) = \frac{\gamma \e{max} a}{K_\gamma + a},
\end{align}
where $\gamma\e{max}$ is the maximal elongation rate and $K_\gamma$ is the energy required for a half-maximal rate. From Eq.~\eqref{eq:trrate} we can compute the total energy consumption by translation of all proteins and get a differential equation for the net turnover of energy:
\begin{equation}
    \dot{a} = n\e{s}v\e{cat} - \sum_{x\epsilon\{r,t,m,q\}}n\e{x}v\e{x} - \lambda a, \label{eq:resourceA}
\end{equation}
where the sum over $x$ is over all types of protein in the cell. Overall, energy is created by metabolizing $s\e{i}$ and lost through translation and dilution by growth. The positive term in Eq.~\eqref{eq:resourceA}, determines energy yield per molecule of internalized nutrient from Eq.~\eqref{eq:nutrient}. The parameter $n\e{s}$ describes the nutrient efficiency of the growth medium.

In rapidly growing \textit{E. coli}, it is known that transcription has a minor role in energy consumption \cite{russell1995energetics}. We therefore model transcription as an energy-dependent process, but with a negligible impact in the overall energy pool. If $w\e{x,max}$ denotes the maximal transcription rate, the effective transcription rate has the form
\begin{equation}
    w\e{x} = w\e{x,max} \frac{a}{\theta\e{x} + a}, \label{eq:wx}
\end{equation}
for all proteins except housekeeping ones, \ie $x\in\{r,t,m\}$. We assume that the transcription of housekeeping mRNAs is subject to negative autoregulation so as to keep constant expression levels in various growth conditions: 
\begin{equation}
   w\e{q} = w\e{q,max} \underbrace{\frac{a}{\theta\e{q} + a}}_{\substack{\text{energy-depedent}\\ \text{translation}}} \times \quad \underbrace{\frac{1}{1+(p_q/K\e{q})^{h\e{q}}}}_{\substack{\text{negative}\\
   \text{autoregulation}}} . \label{eq:wq}
\end{equation}
In Eqs. \eqref{eq:wx} and \eqref{eq:wq},  the parameter $\theta \e{x}$ denotes a  transcriptional threshold, while $K\e{q}$ and $h\e{q}$ are regulatory parameters. The differential equations for the number of mRNAs ($m\e{x}$) 
are therefore:
\begin{equation}
    \dot{m}\e{x} = w\e{x} - (\lambda + d\e{m}) m\e{x} +v\e{x} - k\e{b}p\e{r}m\e{x} + k\e{u}c\e{x}, \label{eq:mRNAs}
\end{equation}
where $x \in\{r,t,m,q\}$. In Eq.~\eqref{eq:mRNAs}, mRNAs are produced through transcription with rate $w\e{x}$, while mRNAs are lost through dilution $\lambda$ and degradation with rate $d\e{m}$. At the same time, mRNAs bind and unbind with ribosomes, so that the ribosome-mRNA complexes ($c\e{x}$)  follow
\begin{equation}
    \dot{c}\e{x} = -\lambda c\e{x} - v\e{x} + k\e{b}p\e{r}m\e{x} - k\e{u}c\e{x}, \label{eq:complexes}
\end{equation}
where $k\e{b}$ and $k\e{u}$ are the rate constants of binding and unbinding. Translation contributes with a positive term to Eq.~\eqref{eq:mRNAs} and a negative term to Eq.~\eqref{eq:complexes}. The differential equations for protein abundance are therefore:
\begin{align}
\dot{p}\e{x} = v\e{x} - \lambda p\e{x},\,\, x\in\{t,m,q\}.
\label{eq:proteome}
\end{align}
We note that Eq.~\eqref{eq:proteome} applies to all proteins except free ribosomes. The equation for free ribosomes $p\e{r}$ includes an additional term:
\begin{equation}
    \dot{p}\e{r} = v\e{r} - \lambda p\e{r} + \sum_{x\in\{r,t,m,q\}}(v\e{x} - k\e{b}p\e{r}m\e{x} + k\e{u}c\e{x}).\label{eq:resourceR}
\end{equation}
Through Eq.~\eqref{eq:resourceR} the model accounts for competition among different mRNAs for free ribosomes, as well as ribosomal autocatalysis. Ribosomal transcripts sequester free ribosomes for their own translation, and the pool of free ribosomes can increase as a result of translation of new ribosomes and, at the same time, the release of ribosomes engaged in translation of non-ribosomal mRNAs.

Finally, it can be shown (details in \cite{weisse2015mechanistic}) that under the assumption of constant average mass, the specific growth rate can be computed in terms of the total number of ribosomes engaged in translation:
\begin{equation}
\lambda = \frac{\gamma(a)}{M}\times \sum_{x\in\{r,t,m,q\}}c\e{x},\label{eq:lambda}
\end{equation}
where $M$ is the constant cell mass. 

Overall, Eqs.~\eqref{eq:nutrient}--\eqref{eq:lambda} constitute the core of the mechanistic growth model. Equations \eqref{eq:resourceA} and \eqref{eq:resourceR}, in particular, model the availability of energy and ribosomes, both regarded as cellular resources shared between metabolism and protein biosynthesis. The model contains 22 parameters. For \emph{E. coli}, some parameter values were mined directly from the literature and others were estimated with Bayesian inference on published growth data \cite{weisse2015mechanistic,scott2010interdependence}. The parameter values are shown in Table \ref{table:2}. We note that we have assumed that all components of the proteome are not subject to active degradation. As we shall see in the next sections, the core model can be extended with gene circuits of varying complexity. 

\linespread{1}
\begin{table}[!h]
\caption{Model parameters for an \emph{Escherichia coli} host, taken from \cite{weisse2015mechanistic}. Units of aa correspond to number of amino acids per cell.} \label{table:2}
\centering
\resizebox{0.78\textwidth}{!}{
\begin{tabular}{lc|c|lc|c}
\hline
\rowcolor{black!10}[\tabcolsep]& parameter & value &  & parameter & value \\ 
&&&&&\\
& $ s $ & 10\textsuperscript{4} (molecules) && $ M $ & 10\textsuperscript{8} (aa) \\
\hline
&&&&&\\
& $ n\e{r} $ & 7459 (aa/molecules)& & $ \theta\e{r} $ & 427 (molecules) \\
\hline
&&&&&\\
& $ \gamma \e{max} $ & 1260 (aa/min molecules)& & $ K_\gamma $ & 7 (molecules) \\
\hline
&&&&&\\
& $ v\e{t} $ & 726 (min\textsuperscript{-1})& & $ K\e{t} $ & 1000 (molecules) \\
\hline
&&&&&\\
& $ v\e{m} $ & 5800 (min\textsuperscript{-1})& & $ K\e{m} $ & 1000 (molecules) \\
\hline
&&&&&\\
& $ w\e{r,max} $ & 930 (molecules / min)& & $ w\e{m,max},\,w\e{t,max} $ & 4.14 (molecules / min) \\
\hline
&&&&&\\
& $ w\e{q,max} $ & 949 (molecules/min)& & $d\e{m}$ & 0.1 (min\textsuperscript{-1}) \\
\hline
&&&&&\\
& $ K\e{q} $ &  152219 (molecules)& & $ h\e{q} $ & 4\\
\hline
&&&&&\\
& $ \theta\e{q},\theta\e{t},\theta\e{m} $ &  4.38 (molecules)& & $ n\e{q},n\e{t},n\e{m} $ & 300 (aa/molecules)\\
\hline
&&&&&\\
& $ k\e{b}$ &   0.0095 ($\text{min}^{-1}\text{molecules}^{-1}$)& & $ k\e{u} $ & 1 (min\textsuperscript{-1})\\

\end{tabular}}
\end{table}

\linespread{1.5}
\section{Modelling gene circuits coupled with their host}\label{sec:host-circuit}
In this section we discuss how to extend the mechanistic growth model with heterologous circuit genes. The extended model can be employed for predicting the impact of genetic parameters, such as promoter strengths or gene length, on the growth rate of the host strain and the resulting heterologous expression levels. We first describe the steps needed to extend the model, and then illustrate the ideas with a simple model for an inducible gene. This is a simple example that contains all the elements needed by more complex circuits.

\subsection{Extending the model with heterologous genes}\label{ssec:extend}
The extension of the model requires three steps:

\paragraph{Step 1: add new model species.} First, we include mass balance equations for the expression of each heterologous gene. This requires three additional species per gene: the transcript, the mRNA-ribosomal complex and the protein, all of which follow dynamics similar to Eqs. \eqref{eq:mRNAs}--\eqref{eq:proteome}:
\begin{align}
\begin{split}
\dot{p}_{i}^{\text{c}} &= v_{i}^{\text{c}} - (\lambda + d\e{p})p_{i}^{\text{c}},\\
\dot{m}_{i}^{\text{c}} &= w_{i}^{\text{c}} - (\lambda + d\e{m}) m_{i}^{\text{c}} +v_{i}^{\text{c}} - k_{\text{b},i\,}^{\text{c}}p\e{r}m_{i}^{\text{c}} + k_{\text{u},i\,}^{\text{c}}c_{i}^{\text{c}},\\
\dot{c}_{i}^{\text{c}} &= -\lambda c_{i}^{\text{c}} + k_{\text{b},i\,}^{\text{c}}p\e{r}m_{i}^{\text{c}} - k_{\text{u},i\,}^{\text{c}}c\e{x} - v_{i}^{\text{c}}, \label{eq:circeq}
\end{split}
\end{align}
where the superscript $\text{c}$ denotes heterologous species and the subscript $i$ denotes the $i^{\text{th}}$ heterologous gene. The ribosomal binding parameters $k_{\text{b},i}^{\text{c}}$ and $k_{\text{u},i}^{\text{c}}$ are specific to each gene and can be used, for example, to model different ribosomal binding sequences. The translation rate $v\e{i}^{\text{c}}$ is modelled similarly as that of native genes in Eq.~\eqref{eq:trrate}:
\begin{align}
v\e{i}^{\text{c}} = \frac{c_{i}^{\text{c}}}{n_{i}^{\text{c}}} \times \frac{\gamma\e{max} a}{a + K_\gamma}, \label{eq:treqi}
\end{align}
with $n\e{i}^{\text{c}}$ being the length of the $i^{\text{th}}$ circuit protein. Likewise, the transcription rate is similar to Eq.~\eqref{eq:wx}:
\begin{align}
w_{i}^{\text{c}} = w_{\text{max},i\,}^{\text{c}} \frac{a}{\theta^{\text{c}} + a} R_i, \label{eq:wheti}
\end{align}
where $w_{\text{max},i\,}^{\text{c}}$ is the maximal transcription rate. Note that we have included an additional term $R_i$ to model regulatory interactions by other genes. Complex circuit connectivities can be modelled by suitable choices of the function $R_i$. Later in Section 4 we exemplify this with models for transcriptional logic gates.

\paragraph{Step 2: modify allocation of resources.} Second, we include the additional consumption of energy and ribosomes in the model. Starting from the resource equations in Eqs. \eqref{eq:resourceA} and \eqref{eq:resourceR}, we write:
\begin{align}
&\dot{a} = n\e{s}v\e{cat} - \sum_{x}n\e{x}v\e{x} - \underbrace{\sum_{i}n_{i}^{\text{c}}v_{i}^{\text{c}}}_{\substack{\text{energy consumption}\\ \text{by foreign genes}}} - \,\,\lambda a \label{eq:resourceA2},\\
&\\
&\dot{p}\e{r} = v\e{r} - \lambda p\e{r} + \sum_{x}(v\e{x} - k\e{b}p\e{r}m\e{x} + k\e{u}c\e{x})\notag + \underbrace{\sum_{i}(v_{i}^{\text{c}} - k_{\text{b},i}^{\text{c}}p\e{r}m_{i}^{\text{c}} + k_{\text{u},i}^{\text{c}}c_{i}^{\text{c}})}_{\substack{\text{consumption of free ribosomes}\\ \text{by foreign genes}}}.\notag
\end{align}

\paragraph{Step 3: adjust growth rate prediction.} Third, we update the prediction of growth rate in Eq.~\eqref{eq:lambda} to include translation of heterologous genes:
\begin{align}
	\lambda &= \frac{\gamma (a)}{M} \bigg( \sum_{x}c\e{x} + \underbrace{\sum_{i}c_{i}^{\text{c}}}_{\substack{\text{ribosomal}\\\text{complexes }}}\bigg).\label{eq:lambda2}
\end{align}

\subsection{Simulation of an inducible gene}
Inducible expression systems are widely employed as building blocks of complex gene circuits. As an example, we consider a reporter gene (\emph{rep}) under the control of an inducible promoter, modelled by the reactions in Table \ref{table:rep}.
\begin{table}[!h]
\caption{Reactions for an inducible reporter gene.\label{table:rep}}
\centering
\resizebox{\textwidth}{!}{%
\begin{tabular}{bababac}
 & transcription & dilution/degradation & ribosome binding & dilution & translation & dilution/degradation\\
 
 \rowcolor{Gray}REP &\cellcolor{LightCyan} $\phi \xrightarrow{w\e{rep}} m\e{rep}$ & $m\e{rep} \xrightarrow{\lambda + d\e{m,rep}} \phi$ &\cellcolor{LightCyan}$p\textsubscript{r} + m\e{rep} \xrightleftharpoons[k\e{u}]{k\e{b}} c\e{rep}$ &  $c\e{rep} \xrightarrow{\lambda} \phi$ &\cellcolor{LightCyan} $n\e{r} a + c\e{rep} \xrightarrow{v\e{rep}} p\e{r} + m\e{rep} + p\e{rep}$&  $p\e{rep} \xrightarrow{\lambda + d\e{p,rep}} \phi$\\ 
\end{tabular}}
\end{table}

The model contains mRNAs of the heterologous gene, which can reversibly bind to free ribosomes of the host, $p\e{r}$. Protein translation consumes energy ($a$) and, at the same time, proteins and other model species are diluted by cell growth. In contrast to native proteins of the host, however, we assume that heterologous proteins are tagged for degradation by proteases, a strategy often employed to accelerate protein turnover \cite{mcginness2006engineering}. This active degradation is modelled by the parameter $d\e{p,rep}$ in Table \ref{table:rep}.

We do not explicitly model the molecular mechanism for induction, as this will depend on the particular implementation of choice. For example, in the \emph{tetR} inducible system, the inducer anhydrotetracycline (aTc) activates gene expression by reversible binding to the tetracyline repressor \emph{tetR}, whereas in the \emph{lac} inducible system, the inducer Isopropyl-$\beta$-D-thiogalactoside (IPTG) binds to allosteric sites of the lac repressor \emph{lacR}. Instead, we lump the induction mechanism into an effective transcription rate, denoted as $w\e{rep}$ in Table \ref{table:rep}.

Using the general circuit equations in  \eqref{eq:circeq}--\eqref{eq:wheti} of Section \ref{ssec:extend}, for the inducible gene Eq.~\eqref{eq:circeq} becomes:
\begin{align}
\begin{split}
\dot{p}\e{rep} &= v\e{rep} - (\lambda + d\e{p,rep})p\e{rep},\\
\dot{m}\e{rep} &= w\e{rep} - (\lambda + d\e{m,rep}) m\e{rep} +v\e{rep} - k\e{b,rep}p\e{r}m\e{rep} + k\e{u,rep}c\e{rep},\\
\dot{c}\e{rep} &= -\lambda c\e{rep} + k\e{b,rep}p\e{r}m\e{rep} - k\e{u,rep}c\e{rep} - v\e{rep}.\label{eq:circeq_h}
\end{split}
\end{align}

\noindent The rate of reporter translation follows as in Eq.~\eqref{eq:treqi}:
\begin{align}
v\e{rep} = \frac{c\e{rep}}{n\e{rep}} \times \frac{\gamma\e{max} a}{a + K_\gamma}, \label{eq:treqi}
\end{align}
where $n\e{rep}$ is the length of the reporter in amino acids. Likewise, the transcription rate in Eq.~\eqref{eq:wheti} becomes:
\begin{align}
w\e{rep} = w\e{max,\text{rep}}\times  \frac{a}{\theta^{\text{c}} + a}.\label{eq:wheti_het}
\end{align}

\noindent Note that in the transcription rate, the regulatory term is $R_i=1$, because the inducible system does not contain any regulatory interactions.

\begin{figure*}[!h]
\centering
\includegraphics[scale=1]{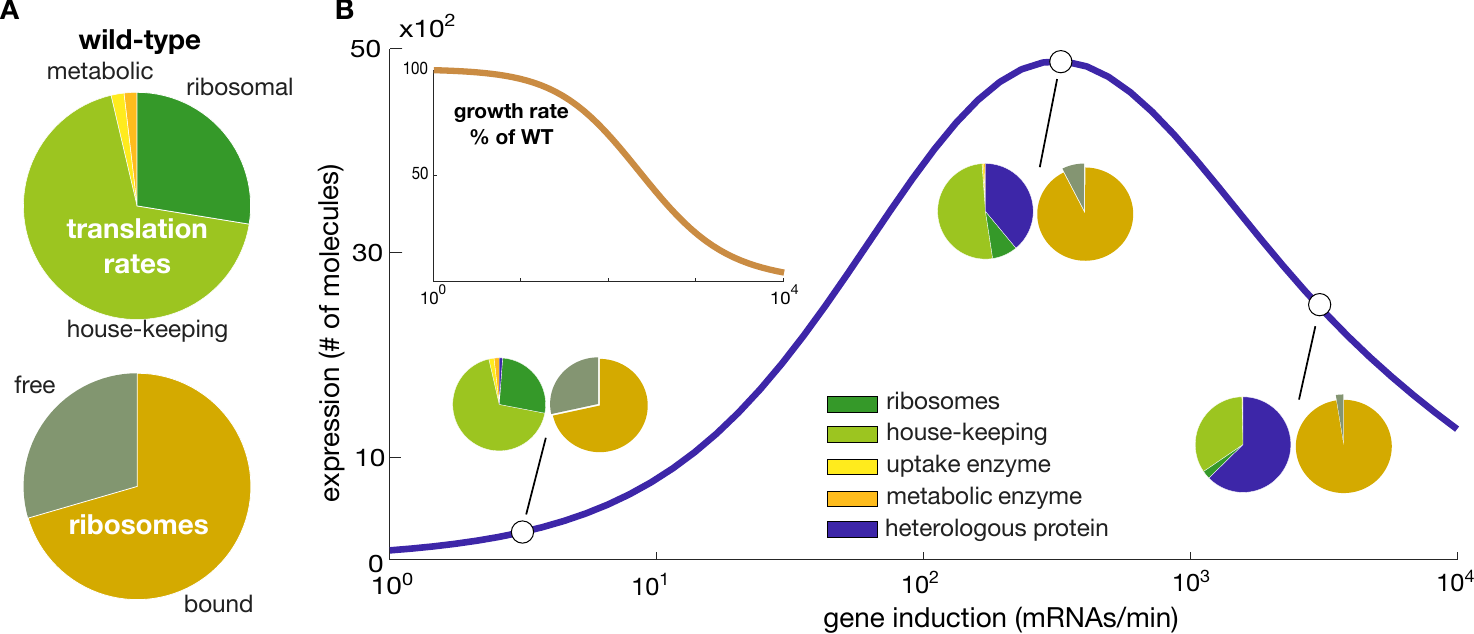}
\caption{\textbf{Simulation of an inducible gene.} \textbf{(A)} Steady state translation rates and ribosomal abundance predicted for the wild-type \emph{Escherichia coli} model, parameterized as in Table \ref{table:2}. \textbf{(B)} Predicted steady state expression of a heterologous gene for increasing induction strength. The pie charts indicate translation rates and ribosomal abundance as in the left panel. The inset shows the predicted growth rate, relative to the wild-type. The induction strength was modelled with the parameter $w\e{max,rep}$ in Eq.~\eqref{eq:wheti_het}. The binding rate constant was set equal to the dissociation rate constant, so that $k\e{b,rep} = 1\times10^{-2}\,\text{min}^{-1}\text{molecules}^{-1}$, $k\e{u,rep} = 1\times10^{-2}\,\text{min}^{-1}$. Transcript and protein half-lives were set to two and four minutes, respectively \cite{elowitz2000synthetic}, so that $d\e{m,rep} = \ln 2\slash2\,\text{min}^{-1}$ and $d\e{p,rep} = \ln 2\slash4\,\text{min}^{-1}$.}
\label{fig3}
\end{figure*}

Before simulating the expression of the heterologous protein, we first need to obtain an estimate for the proteome composition of the wild-type. This is required to initialize the host-circuit simulations with a physiologically realistic cellular composition. To this end, we first simulate Eqs.~\eqref{eq:nutrient}-\eqref{eq:lambda} for the ``wild-type model'' until steady state. The results, summarized in \figref{fig3}A, show that host proteins are translated at different rates with most of the translating ribosomes bound to mRNAs of house-keeping proteins. However, a sizeable fraction is bound to ribosomal mRNA, highlighting how the growth model accounts for ribosomal autocatalysis. A closer look (\figref{fig3}A, bottom) reveals that translation-engaged ribosomes account approximately for two-thirds of the total ribosomal fraction in the form of mRNA-ribosomal complexes, with one-third remaining free.

Next, we simulate heterologous expression using the maximal transcription rate $w\e{max,rep}$ in Eq.~\eqref{eq:wheti_het} to describe the effect of different gene induction strengths. As shown in the dose-response curve in \figref{fig3}B, the model predicts that increased induction causes an increase in expression. We observe, however, that protein expression reaches a maximum at a critical induction strength and subsequently drops sharply for stronger induction. This reflects the limitations that resource competition imposes on the expression of a heterologous gene \cite{nikolados2019growth}.

To understand the main source of the resource limitations, we use the model to explore the synthesis rates of the various components of the proteome. Because growth rate is linearly related to the total rate of translation (Eq.~\eqref{eq:lambda2}), we can make direct conclusions for cellular growth as well. As shown in \figref{fig3}B (inset), the model predicts a sigmoidal decrease in growth rate for stronger gene induction. At low induction, expression of the foreign gene is mostly at the expense house-keeping proteins, while ribosomes, transporter and metabolic enzymes, show little decrease. This suggests that the host can compensate for this load through transcriptional regulation and repartitioning of the proteome (\figref{fig3}B). As the induction of the reporter gene increases, circuit mRNAs dominate the mRNA population, hence increasing the competition for free ribosomes. Finally, for sufficiently strong induction, ribosomal scarcity leads to reduction of all proteins, which in turn leads to the drop in growth rate observed in \figref{fig3}B (inset). These results are in agreement with the widespread conception that ribosomal availability is a major control node for cellular physiology \cite{scott2010interdependence,vind1993synthesis,dong1995gratuitous}, with depletion of free ribosomes being the main source of burden for translation of circuit genes \cite{ceroni2015quantifying,gorochowski2016minimal}.

\section{Simulation of transcriptional logic gates}\label{sec:gates}
There has been substantial interest in gene constructs that mimic digital electronic circuity \cite{hasty2002engineered,lim2010designing,khalil2010synthetic}. Cellular logic gates, in particular, have been used to produce desired behaviours in response to various inputs such as temperature, pH and small molecules \cite{joshi2009novel,paitan2004monitoring,saeidi2011engineering}. Multiple logic gates can be combined to build larger information-processing circuits with advanced cellular functions  \cite{tabor2009synthetic}.

\begin{figure*}[!h]
\centering
\includegraphics{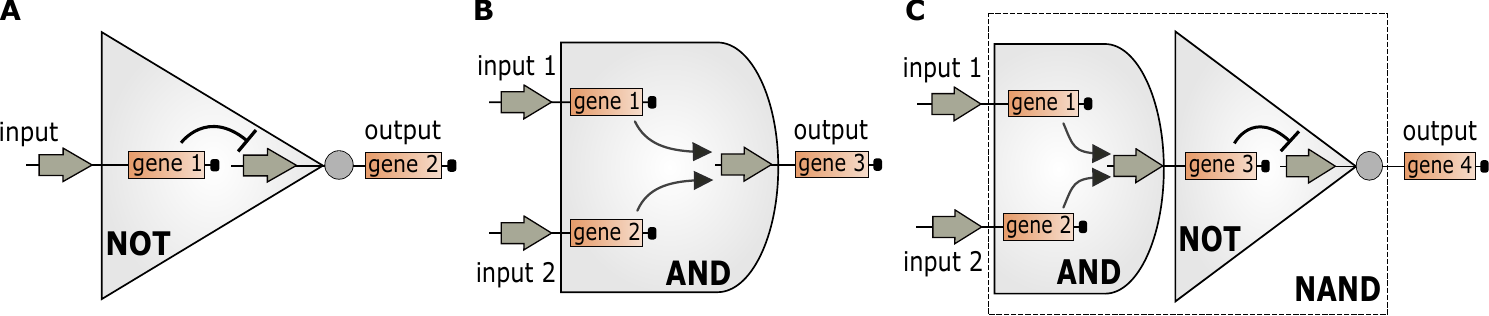}
\caption{\textbf{Logic gates based on transcriptional regulators.} \textbf{(A)} The NOT gate contains two genes connected in cascade. Repression of gene 2 inverts the input signal. \textbf{(B)} The AND gate contains three genes, in which two transcriptional activators jointly trigger the expression of a third output gene. \textbf{(C)} The NAND gate contains four genes and is the composition of an AND and a NOT gate. Circuit connectivities are based on the implementation by Wang \emph{et al} \cite{Wang2011}.}
\label{fig4}
\end{figure*}

To illustrate our simulation strategy in more complex circuitry, here we build host-circuit models for cellular logic gates based on transcriptional regulators \cite{Wang2011}. We first build and simulate the models for a NOT, AND and NAND gates shown in \figref{fig4}. To highlight the power of our approach for circuit design, we then use the host-circuit models to predict circuit function across the design space, using combinations of RBS strength and growth media. As discussed in Section \ref{ssec:extend}, we model the circuits by adding extra genes to the growth model and modifying the mass balance and growth rate equations. We model the circuit connectivity by choosing suitable regulatory terms $R_i$ in the transcription rates in Eq.~\eqref{eq:wheti}, and the gate inputs via the maximal transcription rate $w_{\text{max},i}^{\text{c}}$. 

To compare our host-circuit simulations with those of traditional models, we built circuit-only models using mass balance equations for mRNAs and proteins:
\begin{align}
\begin{split}
    \dot{m}_i^{\text{c}} &= w_i^{\text{c}}R_i - (\lambda^{\text{eff}} + d\e{m})m_i^{\text{c}}, \\
    \dot{p}_i^{\text{c}} &= k^{\text{eff}}_i m_i^{\text{c}} - (\lambda^{\text{eff}} + d\e{p})p_i^{\text{c}},
    \end{split}\label{eq:logic_gates}
\end{align}
where the subscript $i$ denotes the $i^{\text{th}}$ circuit gene and we assume a constant dilution rate, $\lambda^{\text{eff}} =$ 0.022 min\textsuperscript{-1}, which is equal to the growth rate predicted by the model for the wild-type with a nutrient efficiency of $n_s = 0.5$. The effective translation rates are fixed to $k^{\text{eff}}_1=k^{\text{eff}}_2=16.8\,\text{min}^{-1}$  and $k^{\text{eff}}_3=0.61\,\text{min}^{-1}$ for the AND gate, and $k^{\text{eff}}_1=k^{\text{eff}}_2=13.86\,\text{min}^{-1}$,  $k^{\text{eff}}_3=0.058\,\text{min}^{-1}$, and $k^{\text{eff}}_4=347\,\text{min}^{-1}$ for the NAND gate. In all cases, we assume that mRNAs and proteins are actively degraded with rate constants $d_m = \ln 2\slash 2\,\text{min}^{-1} $ and $d_p = \ln 2\slash 4\,\text{min}^{-1}$.
 
\subsection{Host-aware NOT gate}
The NOT gate contains two genes in cascade, where gene 1 codes for a transcriptional repressor that inhibits the expression of gene 2; the circuit diagram is shown in \figref{fig4}A. We first model the NOT gate in isolation using Eq.~\eqref{eq:logic_gates}. We choose the regulatory functions $R_i$ as
\begin{align}
R_1 = 1, \,\, R_2 = \frac{1}{1+\left(\dfrac{p\e{1}^{\text{c}}}{K^{\text{c}}}\right)^{h}}.
\label{eq:regNOT}
\end{align}
The choice of $R_2$ models the inhibition of gene 2, and different inhibitory strengths and cooperativity effects can be described by suitable choices of the threshold $K\e{1}^{c}$ and Hill coefficient $h$. We fix $K\e{1}^{c} = 250$ molecules and $h_1 = 2$.

\begin{figure*}[!h]
\centering
\includegraphics{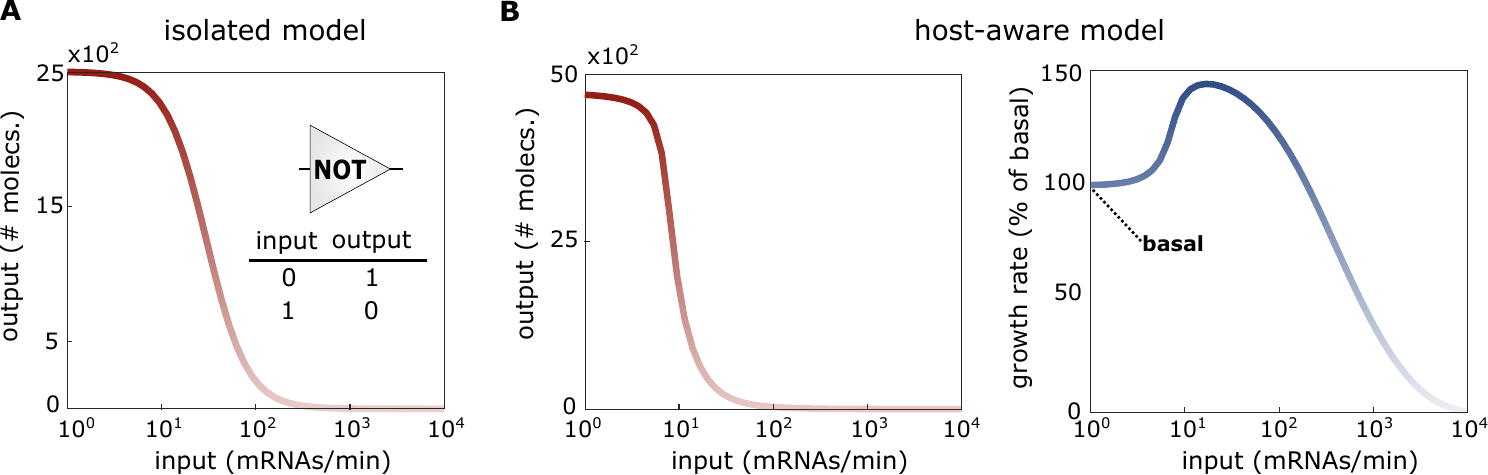}
\caption{\textbf{Host-aware simulation of a NOT gate.} \textbf{(A)} Gate output predicted by a model isolated from the cellular host. Inset shows the boolean truth table for the NOT gate. \textbf{(B)} Output and growth rate predictions from host-aware model of the NOT gate. Growth rate is normalized to a basal case.}
\label{fig5}
\end{figure*}

As shown in \figref{fig5}A, the isolated models correctly predicts the expected circuit function, with stronger induction of the input gene 1 gradually suppressing the expression of the output proteins ($p\e{2}^{c}$), with strong induction resulting in minimal output yield. In other words, the gate has high output only when the input signal is low, in effect acting as an inverter of the input signal.
 
To simulate the host-aware NOT gate, we follow the procedure outlined in Section \ref{ssec:extend}. The host-aware simulations shown in \figref{fig5}B suggest that the function of the NOT gate remains largely unaffected by host-circuit interactions. For intermediate input levels, simulations predict an increase in growth rate of up to $\sim$50\% with respect to a basal case. Such apparent growth benefit is a consequence of the circuit architecture (\figref{fig4}A): an increase in the input causes a stronger repression of gene 2 and thus relieves the burden on the host. But since the expression of the repressor coded by gene 1 also burdens the host, for high inputs the expression of gene 1 counteracts the growth advantages gained by repression of gene 2, resulting in an overall drop in growth rate.

\subsection{Host-aware AND gate} The AND gate comprises two genes that co-activate a third output gene (\figref{fig4}B). As built in the original implementation \cite{Wang2011}, the promoter for gene 3 is activated only when both the co-dependent enhancer-binding proteins, encoded by genes 1 and 2, are present in a heteromeric complex. Consequently, the regulatory functions for the AND gate are:
\begin{align}
R_1 = 1,\,\,  R_2 = 1,\,\,  R_3 = \frac{\left(\dfrac{p\e{1}^{\text{c}}}{K_1^{\text{c}}}\right)^{h_1}}{1+\left(\dfrac{p\e{1}^{\text{c}}}{K_1^{\text{c}}}\right)^{h_1}}\times\dfrac{\left(\dfrac{p\e{2}^{\text{c}}}{K_2^{\text{c}}}\right)^{h_2}}{1+\left(\dfrac{p\e{2}^{\text{c}}}{K_2^{\text{c}}}\right)^{h_2}},
\label{eq:regAND}
\end{align}
with $K\e{1}^{c} =$ 200 molecules and $h\e{1} =$ 2.381 for the activation by gene 1, and $K\e{2}^{c} =$ 3000 molecules and $h\e{2} =$ 1.835 for the activation by gene 2; these values are similar to the parameter values estimated in Wang \emph{et al} \cite{Wang2011}. 

\begin{figure*}[!h]
\centering
\includegraphics{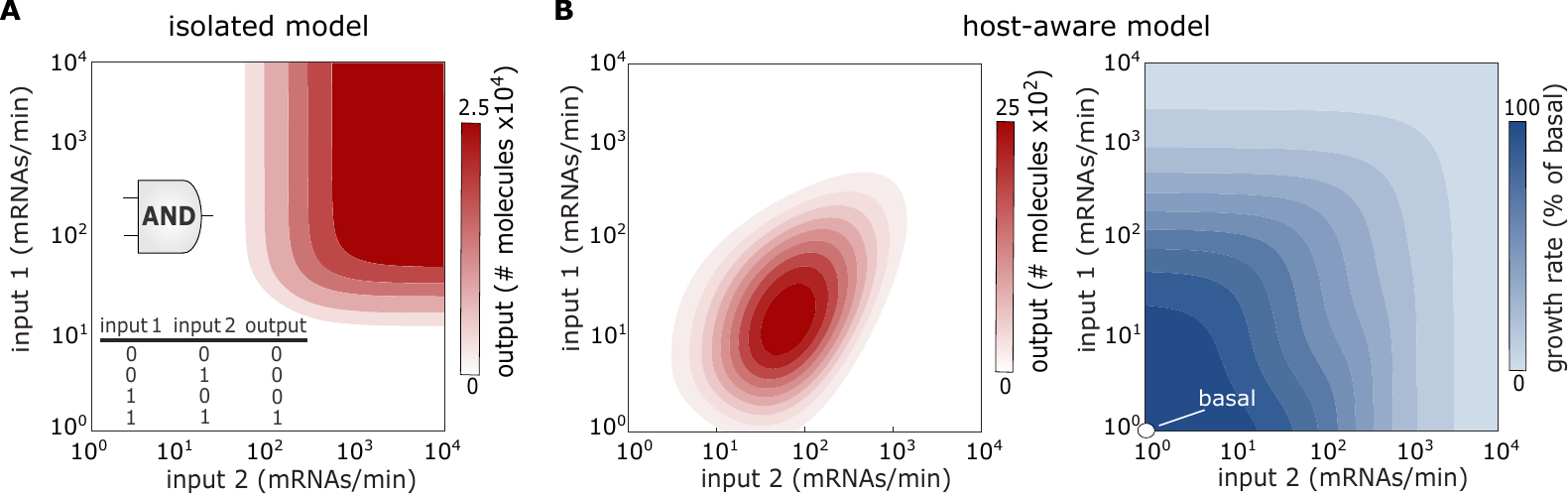}
\caption{\textbf{Host-aware simulation of an AND gate.} \textbf{(A)} Output predicted by a model isolated from the cellular host. Inset shows the boolean truth table for the AND gate. \textbf{(B)} Output and growth rate predictions from host-aware model of the AND gate across the input space. Growth rate is normalized to the basal case in lower left corner of the heatmap.}
\label{fig6}
\end{figure*}

Simulations of the isolated model (\figref{fig6}A) show that, as expected, the gate has a high output only when the input signals are high. This agrees with the expected truth table of the AND, shown in the inset of \figref{fig6}A. In contrast, simulations of the host-aware model, shown \figref{fig6}B, suggest a strong impact of host-circuit interactions. The host-aware model predicts a bell-shaped response surface, where the output reaches a maximal value for an intermediate level of the inputs, beyond which the output drops monotonically. Such loss of function coincides with a drop in growth rate observed for increased levels of either input, as seen in the right panel of \figref{fig6}B, and thus suggests a link between growth defects and poor circuit function.

\subsection{Host-aware NAND gate} The NAND gate is the negation of an AND gate, and thus produces a low output only when both inputs are high. As shown in  \figref{fig4}C, the gate has four genes connected as the composition of an AND and NOT gates. As with the previous two cases, we simulate the isolated model using Eq.~\eqref{eq:logic_gates}. The regulatory functions for the NAND gate are:
\begin{align}
\begin{split}
R_1 &= 1,\\
R_2 &= 1, \\
R_3 &= \frac{\left(\dfrac{p\e{1}^{\text{c}}}{K_1^{\text{c}}}\right)^{h_1}}{1+\left(\dfrac{p\e{1}^{\text{c}}}{K_1^{\text{c}}}\right)^{h_1}}\times\dfrac{\left(\dfrac{p\e{2}^{\text{c}}}{K_2^{\text{c}}}\right)^{h_2}}{1+\left(\dfrac{p\e{2}^{\text{c}}}{K_2^{\text{c}}}\right)^{h_2}},\\
R_4  &= \dfrac{1}{1+\left(\dfrac{p\e{3}^{\text{c}}}{K_3^{\text{c}}}\right)^{h_3}},\end{split}\label{eq:regNAND}
\end{align}
with parameter values for $R_3$ equal to those for $R_3$ of the AND gate in Eq.~\eqref{eq:regAND}, and parameter values for $R_4$ equal to those of $R_2$ for the NOT gate in Eq.~\eqref{eq:regNOT}.

\begin{figure*}[!h]
\centering
\includegraphics{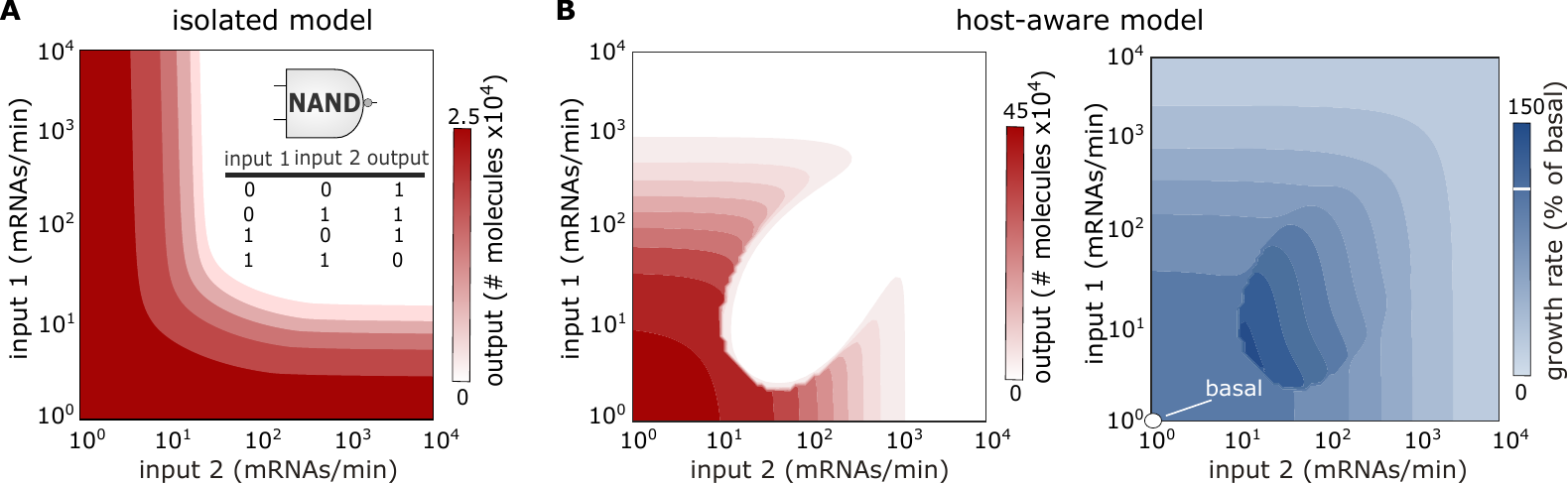}
\caption{\textbf{Host-aware simulation of a NAND gate.} \textbf{(A)} Output predicted by a model isolated from the cellular host. Inset shows the Boolean truth table for the NAND gate. \textbf{(B)} Output and growth rate predictions from host-aware model of the AND gate across the input space. Growth rate is normalized to the basal case in lower left corner of the heatmap.}
\label{fig7}
\end{figure*}

As shown in \figref{fig7}, simulations reveal substantially different predictions between the isolated and host-aware models of the NAND gate. The host-aware model predicts a complex relation between inputs and output that differs from the ideal response predicted by the isolated model. Host-aware simulations produce the correct response across a range of the input space (\figref{fig7}B), but display significant distortions possibly caused by the loss-of-function of the AND component shown in \figref{fig6}B. The impact of host-circuit interactions can also be observed in the predicted growth rate, which suggests a growth advantage for intermediate levels of the inputs. This is a result of the architecture of the NOT gate, akin to what we observed in \figref{fig5}B.

\subsection{Impact of design parameters on circuit function}
In this final section, we conduct a series of simulations that mimic experiments commonly used in circuit design. These aim to explore the impact of design parameters and growth media on circuit function. 
\subsubsection{Ribosomal binding sites (RBS)}
A number of studies have shown that RBS strength is a key modulator of cellular burden \cite{ceroni2015quantifying,gyorgy2015isocost,gorochowski2016minimal,carbonell2015dealing}. Here we examine the impact of RBS strengths on the AND and NAND gates from the previous section. Using the notation in our model, see \eg Eq.~\eqref{eq:circeq}, we define the RBS strength as:
\begin{align}
    \text{RBS}_i &= \frac{k\e{b,i}^{\text{c}}}{k\e{u,i}^{\text{c}}}, \label{eq:RBS}
\end{align}
where $k\e{b,i}^{\text{c}}$ is the mRNA-ribosome binding rate constant (in units of $\text{min}^{-1}\text{molecules}^{-1}$), and $k\e{u,i}^{\text{c}}$ is their dissociation rate constant (in units of $\text{min}^{-1}$). 

\begin{figure*}[!h]
\centering
\includegraphics{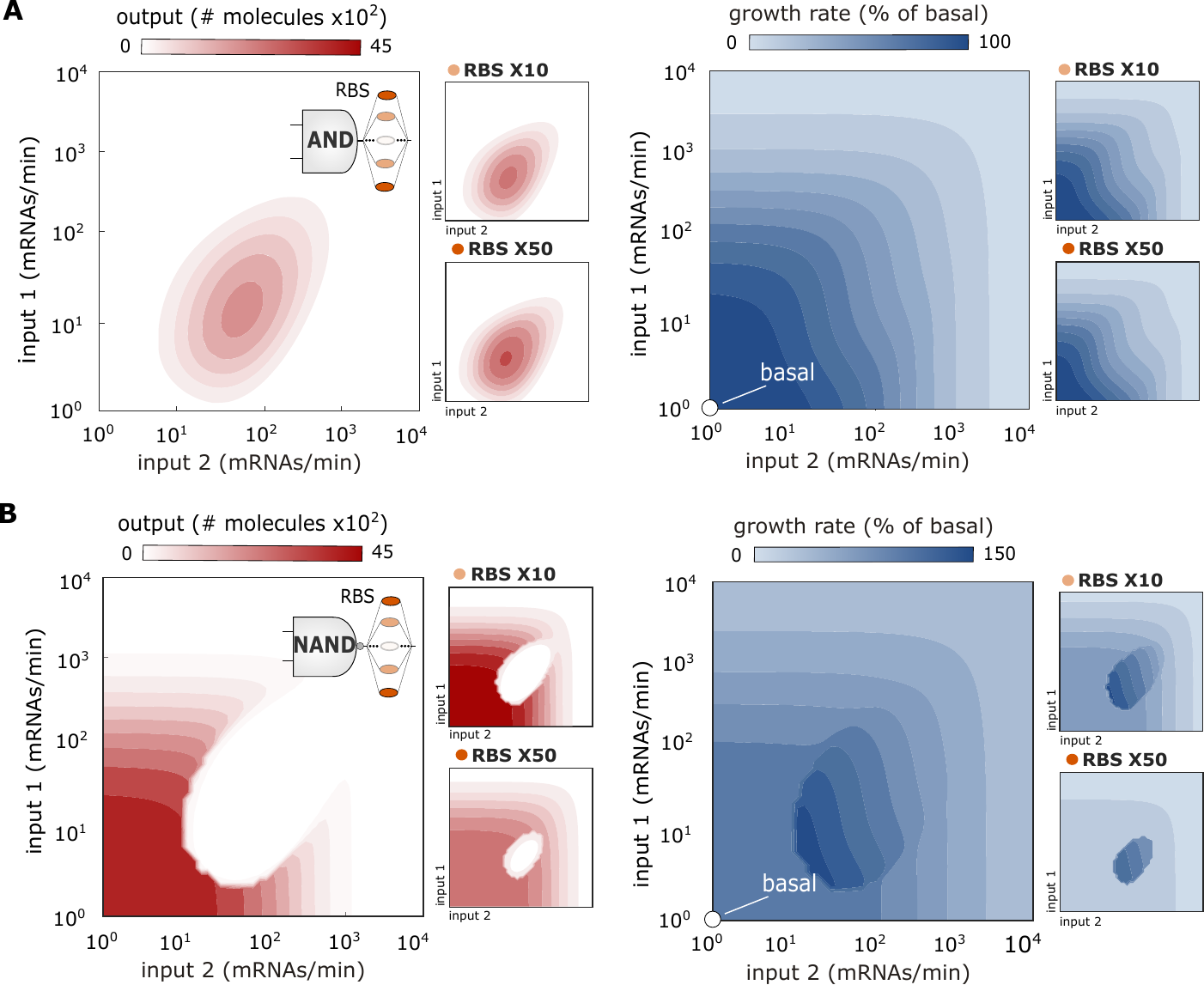}
\caption{\textbf{Impact of ribosomal binding site (RBS) strength.} \textbf{(A)} Output and growth rate predictions for the AND gate in \figref{fig4}B and three RBS strengths. \textbf{(B)} Output and growth rate predictions for the NAND gate in \figref{fig4}C. RBS strengths were computed from Eq.~\eqref{eq:RBS} by simultaneously increasing the binding rate constant $k\e{b,i}^{\text{c}}\in\{10^{-2},10^{-1.5},10^{-1.155}\}$ and decreasing the dissociation rate constant $k\e{u,i}^{\text{c}}\in\{10^{-2},10^{-2.5},10^{-2.855}\}$ in a pairwise manner for $i=3$ (AND gate) and $i=4$ (NAND gate). Gene induction strengths were varied in the range $10^0\leq w\e{max,i}^{c}\leq 10^4$ mRNAs/min for $i=1,2$ in both gates, and fixed $w\e{max,3}^{c} = 375$ mRNAs/min for the AND gate, and $w\e{max,3}^{c} = 375$ mRNAs/min and $w\e{max,4}^{c} = 250$ mRNAs/min for NAND gate.}
\label{fig8}
\end{figure*}

We simulated the AND and NAND gates with variable RBS strengths and gene induction strengths. As shown in \figref{fig8}A (left), the AND gate retains its function for increasing RBS strength. We observe that for the same induction, designs with stronger RBS lead to increased circuit yield. At the same time, the simulations predict (\figref{fig8}A, left) a larger bell-shaped response surface, suggesting, that by increasing RBS, we expect a slightly larger design space where the output can reach a larger maximal value for the same range of inputs. In all cases, however, after the output reaches a maximal value, we find a monotonic drop in circuit yield. The loss-of-function coincides with a drop in growth rate observed in all designs (\figref{fig8}A, right), which becomes more pronounced with stronger RBS.

As shown in \figref{fig8}B, the impact of RBS is more notable for the NAND gate. For designs with stronger RBS (insets \figref{fig8}B, left), but weak induction, the gate displays a behaviour akin to that of the basal case. For intermediate induction, increasing RBS strength has more detrimental effects on the circuit's function. Specifically, the NOT component fails to fully repress the AND component, thus distorting the region where the circuit is functional. However, further increase in RBS, greatly impairs the system leading to near total loss-of-function across the entire response surface (insets \figref{fig8}B, left). Likewise, for stronger RBS and intermediate levels of the input, we observe loss of the growth advantage gained by the NOT gate component (\figref{fig8}B, right).

\subsubsection{Nutrient quality}
Bacterial growth is known to depend critically on the quality of the growth media. As a final illustration of our approach, we used the host-aware models to explore the impact of media on the function of the transcriptional logic gates. We model the quality of the media via the  nutrient efficiency parameter $n_\text{s}$ in Eqs.~\eqref{eq:nutrient} and \eqref{eq:resourceA2}, which determines the energy yield per molecule of internalized nutrient. 

\begin{figure*}[!h]
\centering
\includegraphics{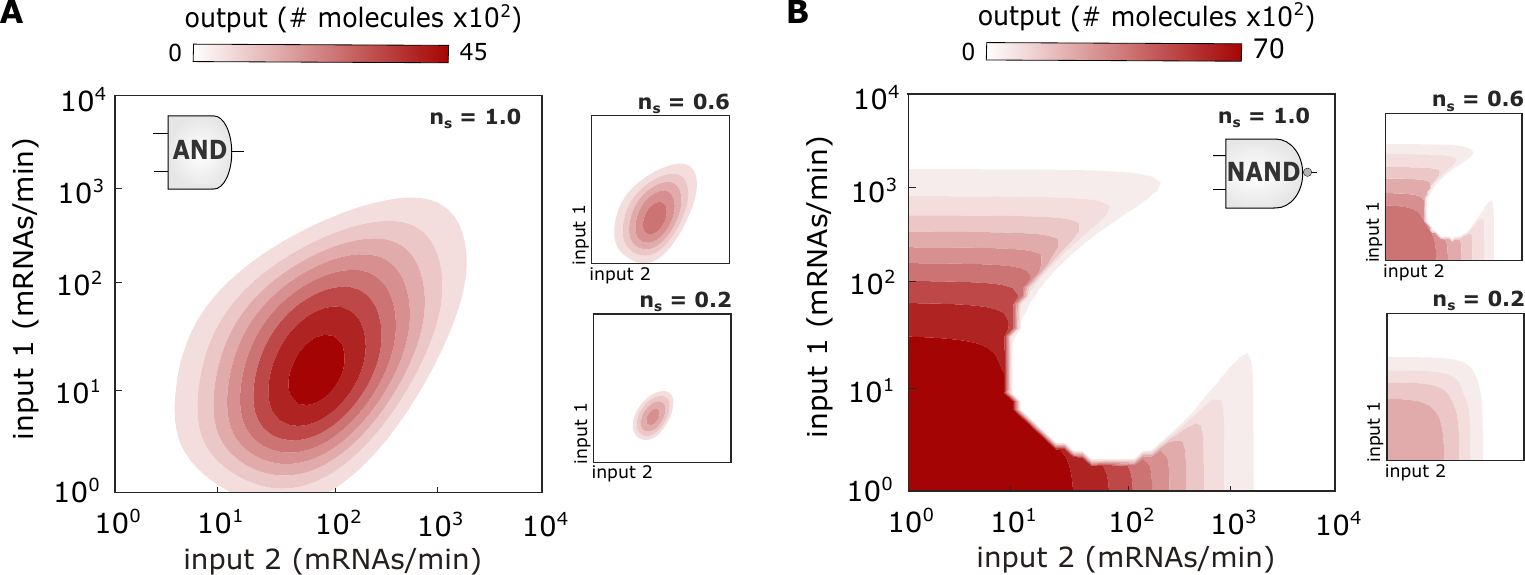}
\caption{\textbf{Impact of growth media on circuit function.} \textbf{(A)} Simulations of the AND gate in \figref{fig4}B in various growth media. \textbf{(B)} Simulations of the NAND gate in \figref{fig4}C in various growth media. In both cases the nutrient quality parameter was set to $n_\text{s}\in\ly0.2,0.6,1.0\ry$; all other model parameters are identical to the simulations in Figures \ref{fig6} and \ref{fig7}B.}
\label{fig9}
\end{figure*}

Our simulations suggest that nutrient quality affects the quantity of output, but not the specific response of the AND gate (\figref{fig9}A). As the quality of the growth medium improves , the gene expression capacity of the host increases and, as a result, we observe an increase operational range of the circuit. However, this is not the case for the NAND gate, which displays a more complex behaviour for low nutrient quality. As seen in \figref{fig9}B, richer media improve the function of the gate, compared to the basal case (\figref{fig7}A). This is because an increase in nutrient quality improves the output of the gate's AND component, which in turn leads to a stronger input for the NOT component, and hence stronger repression. On the contrary, poor nutrient quality leads to loss-of-function for the circuit. As observed in \figref{fig9}A, poorer media correspond to significantly decreased expression of the AND gate, which is also true for the AND component of the NAND gate. This translates to very weak input for the NOT component, which in turn does not properly repress gene 4 (\figref{fig4}C), resulting in the loss of gate functionality (\figref{fig9}B).

\section{Discussion}\label{sec:discussion}

In this chapter we discussed host-aware modelling in Synthetic Biology. Starting from the three bacterial growth laws, we presented a deterministic model to simulate the single-cell dynamics of a bacterial host \cite{weisse2015mechanistic}. We showed how to incorporate synthetic gene circuits into the host model, and used this methodology to simulate host-aware versions of various gene circuits. Finally, we examined the impact of host-circuit interactions on the gates, for combinations of inputs, RBS strength, and growth media of different nutrient quality.

While we focused on host-circuit competition for energy and free ribosomes, in practice gene circuits also consume other components that may become resource bottlenecks, such as RNA polymerases and $\sigma$-factors for transcription, or amino acids and tRNAs for translation. Molecular species associated with these processes can be readily incorporated into the growth model. For instance, instead of a single energy resource $a$, the catabolism of the internalised nutrient $s_i$ by the metabolic protein $p_m$, could also produce a pool of amino acids, which would then participate in the downstream transcription and translation processes. Explicit models of amino acid pools could be employed to study amino acid recycling after protein degradation, or global effects such as upregulation of transcription triggered by nutrient starvation \cite{liao2017integrative,Hartline2020}. Such extensions, however, need to be dealt with caution since they can increase model complexity, and ultimately obscure the relations between different sources of burden. 

A grand goal of Synthetic Biology is to produce target phenotypes through rational design of gene circuits. As with other engineering disciplines, predictive models are an essential step to accelerate the design cycle, yet current models in synthetic biology are largely under-powered for this task. Integrated host-circuit models can effectively bridge this gap and offer a flexible framework to account for a wide range of resource bottlenecks. For example, recent data \cite{cambray2018evaluation,Borkowski2018} suggest highly nonlinear relations between growth rate and heterologous expression and a sizeable burden caused by metabolic imbalances typically found in pathway engineering \cite{Liu2018}. Such findings raise compelling prospects for the integration of mechanistic cell models with large-scale characterization data, ultimately paving the way for more robust and predictable Synthetic Biology.

\end{document}